\documentclass[floatfix,twocolumn,prl]{revtex4}
\usepackage{graphicx}
\usepackage{amsmath}
\usepackage{amssymb}
\usepackage{bm}
\usepackage{dcolumn}
\usepackage{braket}
\usepackage{color}

\begin{document}

\title{The role of spin-flip transitions in the anomalous Hall effect of FePt alloy}

\author{Hongbin~Zhang}
\author{Frank Freimuth}
\author{Stefan~Bl\"ugel}
\author{Yuriy~Mokrousov}
\email[corresp.\ author: ]{y.mokrousov@fz-juelich.de}
\affiliation{Peter Gr\"unberg Institut and Institute for Advanced Simulation, 
Forschungszentrum J\"ulich and JARA, D-52425 J\"ulich, Germany}
\author{Ivo~Souza}
\affiliation{Centro de F\'{\i}sica de Materiales and DIPC, Universidad del
Pa\'\i s Vasco, 20018 San Sebasti\'an, Spain}
\affiliation{Ikerbasque, Basque Foundation for Science, E-48011 Bilbao, Spain}

\date{\today}

\def\sigpar{\sigma_m}
\def\sigperp{\sigma_\theta}
\def\sigout{\sigma_z}
\def\sigin{\sigma_x}
\def\alphapt{\alpha^{\rm Pt}}
\def\alphapd{\alpha^{\rm Pd}}
\def\alphafe{\alpha^{\rm Fe}}
\def\uu{\upuparrows}
\def\ud{\uparrow\hspace{-0.055cm}\downarrow}
\def\red#1{{\color{red}#1}}

\begin{abstract}

  We carry out {\it ab initio} calculations which demonstrate the
  importance of the non-spin-conserving part of the spin-orbit
  interaction for the intrinsic anomalous Hall conductivity of ordered
  FePt alloys. The impact of this interaction is strongly reduced if
  Pt is replaced by the lighter isoelectronic element Pd.  An analysis
  of the interband transitions responsible for the anomalous velocity
  reveals that spin-flip transitions occur not only at avoided band
  crossings near the Fermi level, but also between well-separated
  pairs of bands with similar dispersions.  We also predict a strong
  anisotropy in the anomalous Hall conductivity of FePt caused
  entirely by low-frequency spin-flip transitions.

\end{abstract}

\maketitle

The intrinsic anomalous Hall effect (AHE)~\cite{Nagaosa:review} and
  spin Hall effect (SHE)~\cite{Hirsch} in solids arise from the opposite
  anomalous velocities experienced by spin-up and spin-down electrons as
  they move through the spin-orbit-coupled bands under an applied
  electric field. In paramagnets, where the bands are spin-degenerate,
  these counter-propagating transverse currents result in a time-reversal
  conserving pure spin current. In ferromagnets, where the bands are
  split by the exchange interaction, the same process generates a net
  time-odd charge current.

The above picture is intuitively appealing, and often leads to correct
  conclusions.  However, it leaves out the fact that in the presence of
  the spin-orbit interaction (SOI) the spin projection along the
  quantization axis is not a good quantum number. This is a particularly
  subtle point regarding the SHE, as the proper definition of the spin
  current becomes problematic when spin is not a conserved
  quantity~\cite{Shi}. More generally, processes which do not conserve
  spin (we shall refer to them as spin-flip processes) are known to play a 
  role in phenomena such as spin relaxation~\cite{Zutic} and 
  magnetocrystalline anisotropy~\cite{MAE}.

How does the lack of spin conservation affect the AHE? 
  To analyze this issue, we begin by noting that the anomalous
  velocity results from virtual interband transitions, and that the
  matrix elements involved are the same which describe magnetic
  circular dichroism (see Eq.~(1) below). In a perturbative expansion
  in powers of the spin-orbit coupling strength, the spin-conserving
  (spin-non-conserving) part of the SOI contributes to the dichroic
  conductivity at first (second) order~\cite{Cooper}.  The effect of
  spin-flip transitions is therefore expected to be comparatively
  small, as confirmed by recent tight-binding calculations of the
  anomalous Hall conductivity (AHC) for the $3d$ transition
  metals~\cite{Kontani}.

It should be kept in mind, however, that in materials containing 
  heavy atoms the SOI cannot be treated as a small perturbation.  
  Moreover, first-principles calculations of the AHC~\cite{Yao} 
  have established the crucial role of near degeneracies across 
  the Fermi level, for which the above arguments, based on
  non-degenerate perturbation theory, do not apply. While the full 
  SOI was included in previous calculations~\cite{Nagaosa:review}, 
  the specific role of the non-spin-conserving part was not thoroughly 
  investigated.
  
In this Letter, we use first-principles calculations to study the
  impact of spin-flip transitions on the intrinsic AHC of FePt orderered
  alloys~\cite{Chen,Seemann}.  
  This material has a number of desirable properties for the 
  present study.  Firstly, the heavy element Pt provides the strong 
  SOI, which can be "tuned" by replacing Pt with Pd. 
  Secondly, previous work has established that in samples with finite
  disorder the intrinsic contribution to the AHC is much larger than
  the extrinsic one~\cite{Seemann}. It is becoming increasingly clear 
  that the AHE in moderately resistive samples of itinerant ferromagnets 
  such as FePt is often dominated by the intrinsic contribution~\cite{Nagaosa:review}, 
  which at present is the only one that can be reliably calculated from 
  first-principles in materials with unknown structural disorder and 
  impurity content. We shall therefore focus exclusively on the intrinsic 
  part of the AHC, neglecting extrinsic contributions such as skew-scattering 
  and side-jump. We find that the contribution of spin-flip transitions to the 
  AHC of FePt is considerable, amounting to about one fifth of the 
  total value. More importantly, the calculations reveal a clear 
  experimental signature of spin-flip transitions: as the magnetization 
  is rotated from the uniaxial direction to the basal plane, their 
  contribution to the AHC changes sign, leading to a factor-of-two 
  reduction in the net AHC. In contrast, the spin-conserving part 
  is almost perfectly isotropic.

We identify two distinct mechanisms for the spin-flip transitions. The
  first involves spin-orbit-induced avoided crossings at the intersections 
  between exchange-split up- and down-spin Fermi-surface sheets 
  (these intersections occur along lines in $k$-space which we
  shall refer to as {\it hot loops}, in analogy with the {\it hot spots}
  which have been discussed in connection with spin-relaxation in
  nonmagnetic metals~\cite{Zutic}).  The second mechanism involves
  spin-orbit driven transitions between bands with similar dispersion
  which are split in energy across the Fermi level. We shall refer to
  them as {\it ladder transitions}. Both occur at low frequencies, of 
  the order of the spin-orbit coupling strength.

\begin{table}[t!]
\begin{ruledtabular}
\begin{tabular}{ccccccc}
    &   & $\sigma^{\rm tot}$ & $\sigma^{\uu}$ & $\sigma^{\ud}$ & $\Delta\sigma^{\uu}$
    & $\Delta\sigma^{\ud}$ \\ \hline
    FePt & $[001]$ & 818.1 &  576.6 & 133.4    &  $-$8.5  &  317.3  \\
         & $[100]$ & 409.5 &  585.1 & $-$183.9 &          &         \\ \hline
    FePd  & $[001]$ & 135.1 &  108.4 &  28.4    & $-$88.5  & $-$33.6 \\
          & $[100]$ & 275.9 &  196.9 &  62.0    &          &         \\
\end{tabular}
\end{ruledtabular}
\label{table:anisotropy}
\caption{
  Values of the AHC in FePt and FePd with the magnetization along
  [001] ($\sigout$) and [100] ($\sigin$). For each orientation,
  $\sigma^\uu$ ($\sigma^{\ud}$) is calculated by keeping only the
  first (second) term in the spin-orbit Hamiltonian (2), while both terms
  are kept when calculating of $\sigma^{\rm tot}$.
  $\Delta\sigma^{\uu(\ud)}$ is defined as the difference between the
  spin-conserving (spin-flip) parts of $\sigma_z$ and $\sigma_x$. All
  values are in S/cm.}
\end{table}

Let us briefly review the formalism for calculating the intrinsic
  AHC from first-principles.  For a ferromagnet with the orthorhombic
  crystal structure and magnetization $\mathbf{M}$ along the $\hat{z}$
  ([001]) axis, the AHC $\sigma_{z}\equiv\sigma_{xy}$ is
  given by the $k$-space integral of the Berry
  curvature~\cite{Yao,Nagaosa:review}:
  \begin{equation}
   \sigma_{z} = \frac{e^2\hbar}{4\pi^3}{\rm Im}\int_{\rm BZ} 
   d\mathbf{k} \sum_{n,m}^{o,e}
   \frac{\Braket{\psi_{n\mathbf{k}}|v_x|\psi_{m\mathbf{k}}}
   \Braket{\psi_{m\mathbf{k}}|v_y|\psi_{n\mathbf{k}}}}
   {(\varepsilon_{m\mathbf{k}}-\varepsilon_{n\mathbf{k}})^2}.
  \end{equation}
  In this expression, $\psi_{n\mathbf{k}}$ and $\psi_{m\mathbf{k}}$
  are respectively the occupied ($o$) and empty ($e$) one-electron 
  spinor Bloch eigenstates of the crystal with eigenvalues 
  $\varepsilon_{n\mathbf{k}}$ and $\varepsilon_{m\mathbf{k}}$, 
  $v_x$ and $v_y$ are Cartesian 
  components of the velocity operator, and the integral is over the 
  Brillouin zone (BZ).  When the direction of $\mathbf{M}$ is changed 
  from the $\hat{z}$-axis to the $\hat{x}$-axis ([100]), the 
  $\sigma_{x}\equiv\sigma_{yz}$ component of the conductivity tensor 
  should be calculated instead, by replacing $v_x\rightarrow v_y$ and 
  $v_y\rightarrow v_z$ in Eq.~(1). 

The calculations were done using the approach of Ref.~\cite{Souza:2006},
  whereby the linear-response expression (1) is rewritten in the basis of
  Wannier functions spanning the occupied and low-lying empty states.
  In this way the infinite sums over bands are replaced by sums over the
  small number of Wannier-interpolated bands.  The Wannier functions
  were generated with {\tt WANNIER90}~\cite{wannier} using the same
  parameters as in Ref.~\cite{Seemann}, by post-processing first-principles
  calculations done using the J\"ulich density-functional theory FLAPW code 
  {\tt FLEUR}~\cite{fleur} (see Ref.~\cite{FLAPW-MLWFs} for details). 
  The unit cell of FePt and FePd contained two atoms in the $L1_0$ structure, 
  with stacking along the [001]-direction. We used the generalized gradient 
  approximation lattice constants of $a=5.14$~bohr~and $c=7.15$~bohr~for 
  FePt, and $a=5.12$~bohr~and~$c=7.15$~bohr~for FePd.

The atomic spin-orbit term in the Hamiltonian has the form
  \begin{equation}
  \label{eq:soi}
    \xi\mathbf{L\cdot S}=
    \xi{\rm L}_{\hat{n}}{\rm S}_{\hat{n}} + \xi\left( 
    {\rm L}^{+}_{\hat{n}}{\rm S}^{-}_{\hat{n}} + 
    {\rm L}^{-}_{\hat{n}}{\rm S}^{+}_{\hat{n}} \right)/2,
  \end{equation}
  where $\xi$ is the spin-orbit coupling strength, $\hat{n}$ is the
  spin magnetization direction (which is taken as the spin-quantization
  axis), $\mathbf{L}$ and $\mathbf{S}$ are the
  orbital and spin angular momentum operators, ${\rm
  L}_{\hat{n}}=\mathbf{L}\cdot\hat{n}$, and ${\rm
  L}^{+}_{\hat{n}}$ and ${\rm L}^{-}_{\hat{n}}$ are the
  corresponding raising and lowering operators (analogously for spin). 
  We shall refer to the first and second terms in Eq.~(2) as the 
  spin-conserving (${\rm LS}^{\uu}$) and spin-flip (${\rm LS}^{\ud}$) 
  parts of the SOI. This terminology refers to the effect of acting with 
  each of them on an eigenstate of $S_{\hat{n}}$. Accordingly, we define
  $\sigma^{\uu}$ and $\sigma^{\ud}$ as the AHC calculated from
  Eq.~(1) after selectively removing ${\rm LS}^{\ud}$ or ${\rm
  LS}^{\uu}$ from Eq.~(2).  This is not an exact decomposition,
  but inspection of Table~I shows that it is approximately valid, as
  $\sigma^{\rm tot} \approx \sigma^{\uu}+\sigma^{\ud}$.

\begin{figure}[t]
\begin{center}
\includegraphics[scale=0.45]{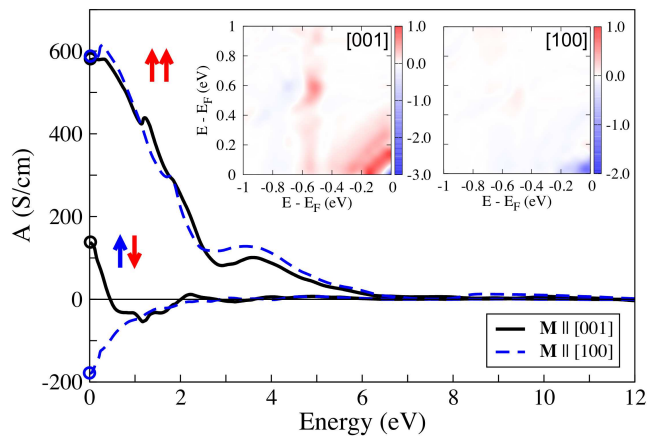}
\end{center}
\caption{\label{fig:cumulative-ahc} (color online)
  Cumulative contribution to the AHC of FePt from the spin-flip ($\ud$) 
  and spin-conserving ($\uu$) dichroic spectra above energy $\omega$.  
  The values of AHC from Table I are indicated as open circles. The two 
  insets display $\Sigma^{\ud}(E_v,E_c)$, the energy-energy density 
  of contributions to $\sigma^{\ud}$, for $\mathbf{M}$ along [001] 
  and [100] (in $10^5$~a.u./${\rm eV}^2$). }
\end{figure}

The importance of spin-flip transitions for the AHC of FePt can be
  seen by analyzing its dependence on the magnetization direction
  (Table~I).  If only the spin-conserving term in Eq.~(\ref{eq:soi}) is
  kept, the resulting AHC $\sigma^{\uu}$ changes by less than 2\% from
  an average value of about 580~S/cm as ${\bf M}$ is tilted from the
  $\hat{z}$-axis to the $\hat{x}$-axis. When the spin-flip term is also
  included, the AHC $\sigma^{\rm tot}$ becomes highly anisotropic,
  decreasing by a factor of two, or roughly 400~S/cm, between [001] and
  [100]. Keeping only the spin-flip part of the SOI reveals that it is
  indeed chiefly responsible for the large anisotropy, as $\sigma^{\ud}$ 
  changes by more than 300~S/cm, from a positive value along [001] to 
  a large negative value along [100].  Such significant AHC anisotropy 
  can occur in uniaxial ferromagnetic crystals, and was previously found 
  in hcp Co~\cite{Souza:2009}. However, in hcp Co the anisotropy is 
  caused by spin-conserving processes. This is also the case for FePd, 
  as seen in Table~I, and we shall comment further on this point below.
  
The AHC can be resolved in energy by defining a {\it cumulative} AHC
  $A(\omega)$, which accumulates all transitions in Eq.~(1) for which
  $\varepsilon_{m\mathbf{k}}-\varepsilon_{n\mathbf{k}}>\omega$
  ~\cite{Souza:2009}. In the limit $\omega\rightarrow 0$ all interband 
  transitions in Eq.~(1) are accounted for, and therefore 
  $A(\omega\rightarrow 0)$ equals the full intrinsic AHC. The spin-conserving 
  and spin-flip cumulative AHCs are plotted in Fig.~\ref{fig:cumulative-ahc} 
  in the range $0\leq\omega\leq 12$~eV, for both $\mathbf{M}\Vert\hat{z}$ 
  and $\mathbf{M}\Vert\hat{x}$. While $A^{\uu}(\omega)$ remains largely
  isotropic over the entire energy range and decays rather slowly with
  $\omega$ up to 4$-$5~eV in energy, $A^{\ud}(\omega)$ picks up only
  for $\omega$ below 1~eV and immediately becomes strongly anisotropic
  with decreasing energy, displaying a characteristic bifurcation
  shape~\cite{Souza:2009}.  Thus, the anisotropy in the AHC arises from
  spin-flip transitions within an energy window of about 0.5~eV around
  $E_F$.

 \begin{table}[t!]
 \begin{ruledtabular}
 \begin{tabular}{ccccccc}
     & ${\rm Fe}^{\rm tot}$  & ${\rm Fe}^{\uu}$ &  ${\rm Fe}^{\ud}$ &
       ${\rm Pt}^{\rm tot}$  & ${\rm Pt}^{\uu}$ &  ${\rm Pt}^{\ud}$ \\ \hline
     $[001]$ & $-$13.7    & 17.9 &  $-$26.8  &  848.0      &  541.0  &  282.3  \\
     $[100]$ &   210.0    & 253.6 & $-$37.5  &  65.0      &  425.7  & $-$360.6   \\
 \end{tabular}
 \end{ruledtabular}
 \label{Table3}
 \caption{
  AHC in FePt for two magnetization directions, resolved into spin-flip
  and spin-conserving contributions from the SOI on each atomic
  species. All values are in S/cm.}
 \end{table}

In order to get further insight into the energy distribution of the spin-flip 
  transitions, we define a new quantity $\Sigma^{\ud}(E_v,E_c)$ as the 
  contribution to $\sigma^{\ud}$ from vertical transitions between pairs 
  of states with energies in the vicinity of $E_v<E_F$ and $E_c>E_F$. 
  Thus, $\iint\Sigma^{\ud}(E_v,E_c) dE_vdE_c=\sigma^{\ud}$, and 
  if the region of integration is restricted to $E_c-E_v>\omega$, we 
  obtain $A^{\ud}(\omega)$.

The function $\Sigma^{\ud}(E_v,E_c)$ is shown in the insets of Fig.~1 
  for the two magnetization directions. In both cases one can see intense 
  blue dots near the origin. They denote large negative contributions 
  concentrated at very low energies, arising from spin-orbit-induced 
  avoided crossings between up- and down-spin Fermi-surface sheets.
  While for $\mathbf{M}\Vert\hat{x}$ these hot-loop features are
  dominant, for $\mathbf{M}\Vert\hat{z}$ a competing positive
  contribution can be clearly seen. It consists of a series of stripes 
  $E_c-E_v\approx\rm{const.}$, with the constant ranging from 0.1
  to 0.5~eV. By analyzing the band structure we find that, owing to
  the off-diagonality of the ${\rm LS}^{\ud}$-operator in the basis of
  localized $d$-orbitals, these transitions come from pairs of bands
  of different orbital character with similar dispersion on either side 
  of $E_F$. Such ladder transitions, indicated schematically in the inset 
  of Fig.~2, provide a different source of AHC. Compared to the hot loops, 
  they do not require band crossings at the Fermi energy, and occur over 
  wider ranges of energy and larger regions of $k$-space. 
  In FePt with
  $\mathbf{M}\Vert\hat{z}$ their contribution is so large that it wins
  over the hot-loop part and determines the sign and magnitude of
  $\sigma^{\ud}$.

The spin-flip processes in FePt are induced mostly by the strong SOI 
  on the Pt atoms. In order to prove this point, we have selectively turned 
  off the SOI on each atomic species inside the crystal. The atom-resolved 
  spin-orbit Hamiltonian reads
  \begin{equation}
    H^{\rm SO} =  \xi_{\rm Fe}\mathbf{L}^{\rm Fe}\cdot\mathbf{S}+
    \xi_{\rm Pt}\mathbf{L}^{\rm Pt}\cdot\mathbf{S},
  \end{equation}
  where $\mathbf{L}^{\mu}$ is the orbital angular momentum operator
  associated with atomic species $\mu$, and $\xi_{\mu}$ is the spin-orbit 
  coupling strength averaged over valence $d$-orbitals. In FePt we find 
  $\xi_{\rm Fe}^0=0.06$~eV and $\xi_{\rm Pt}^0=0.54$~eV, where 
  $\xi^0_\mu$ denotes the value calculated from first-principles.

We have recalculated the AHC after setting to zero either $\xi_{\rm
  Fe}$ or $\xi_{\rm Pt}$ in Eq.~(3), and then using Eq.~(2) to further 
  decompose the remaining term. The results are presented in Table~II. 
  Although such a decomposition is not exact, it reproduces the results of 
  Table I rather well. Namely, the sum of the total conductivities driven by 
  SOI on Fe (Fe$^{\rm tot}$ in Table II) and on Pt (Pt$^{\rm tot}$ in 
  Table II) is in reasonable agreement to the  values of $\sigma^{\rm tot}$ 
  from Table I for both magnetization directions. Moreover, the decomposition 
  of the total atom-resolved AHCs into spin-conserving and spin-flip parts 
  is almost exact, as can be seen from Table II.

  \begin{figure}[t!]
  \begin{center}
  \includegraphics[scale=0.42]{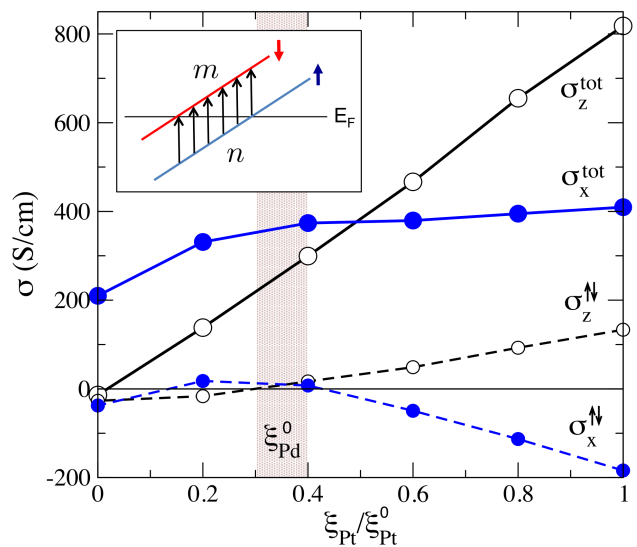}
  \end{center}
  \caption{ (color online) 
  Dependence of the total ($\sigout^{\rm tot}$ and $\sigin^{\rm tot}$) 
  and spin-flip ($\sigout^{\ud}$ and $\sigin^{\ud}$) AHC in FePt alloy 
  on the strength $\xi_{\rm Pt}$ of the SOI inside the Pt atoms.
  The inset depicts schematically the "ladder-type'' spin-flip interband 
  transitions.}
  \end{figure}

Consider first the left part of the Table II,
  where the AHC is driven by $\xi_{\rm Fe}$. For both magnetization
  directions the spin-flip contribution is very small, while the
  spin-conserving part is small along [001] but large along [100]. As
  for the AHC induced by $\xi_{\rm Pt}$, shown on the right-side of
  the table, the spin-conserving part is large but fairly isotropic,
  while the spin-flip part is highly anisotropic, changing from a
  large positive value along [001] to a large negative value along
  [100]. This confirms that the large and strongly anisotropic
  $\sigma^{\ud}$ is governed by the SOI inside the Pt atoms.

A large spin-flip contribution to the AHC in materials with strong
  spin-orbit coupling is perhaps not surprising, given that spin-flip
  transitions appear at second order in a perturbative treatement of the
  SOI. This is confirmed by nonperturbative calculations where we tune
  by hand the SOI strength $\xi_{\rm Pt}$ on the Pt atoms.
  The results for the total and spin-flip AHC are shown
  in Fig.~2 as a function of $\xi_{\rm Pt}/\xi_{\rm Pt}^0$. It can be 
  seen that for $\xi_{\rm Pt}$ less than $\xi_{\rm Pt}^0/2$, the 
  absolute value of the spin-flip AHC does not exceed a modest value 
  of 50~S/cm. In this regime $\sigout^{\rm tot}$ and $\sigin^{\rm tot}$ 
  are dominated by spin-conserving processes.  Moreover, we note that
  while the decrease in $\sigout^{\rm tot}$ is almost perfectly linear, 
  $\sigin^{\rm tot}$ stays fairly constant over a wide region of 
  $\xi_{\rm Pt}$ values. This can be understood from the fact that for 
  $\mathbf{M}\Vert\hat{z}$ the spin-conserving and spin-flip 
  contributions arising from $\xi_{\rm Pt}$ largely cancel one another 
  (see Table~II), so that the total AHC is mostly driven by the SOI on 
  the Fe atoms. In contrast, for $\mathbf{M}\Vert\hat{z}$ it is the
  SOI on the Pt atoms which dictates the AHC.

The artificial tuning of $\xi_{\rm Pt}$ performed above describes rather 
  well what happens if the Pt atoms are replaced with Pd, to form the 
  experimentally known FePd alloy~\cite{Seemann}.  This can be seen 
  by comparing the values of $\sigma^{\rm tot}$ and $\sigma^{\ud}$ 
  for FePd in Table~I with the values taken from the shaded area in Fig.~2, 
  where $\xi_{\rm Pt}\approx \xi_{\rm Pd}^0=0.19$~eV.  In particular, 
  the sign of the AHC anisotropy in FePd, which is opposite from that in FePt, 
  is correctly reproduced by the scaled calculations on FePt.

In summary, we predict a large contribution from spin-flip transitions 
  to the intrinsic AHE of FePt ordered alloys. Such transitions are induced 
  by the strong spin-orbit interaction on the Pt atoms. They are concentrated 
  at frequencies below the spin-orbit interaction energy, and their sign 
  depends on the magnetization direction, making the AHE in this material 
  strongly anisotropic. Our calculations have assumed perfectly ordereded 
  alloys, therefore neglecting extrinsic contributions to the AHC. 
  First-principles methods capable of incorporating the effects of disorder in 
  the calculation of the AHC have been recently developed [16-18]. An 
  interesting direction for future work would be to use such methods to 
  investigate both the role of spin-flip transitions and the orientation 
  dependence of the extrinsic AHE.

We acknowledge discussions with M. Le\v{z}ai\'c, Ph. Mavropoulos and
  K. M. Seemann. This work was supported by the HGF-YIG Programme 
  VH-NG-513 and NSF Grant DMR-0706493. Computational 
  time on JUROPA and JUGENE supercomputers was provided by J\"ulich
  Supercomputing Centre.

\end{document}